\documentclass[11pt,twoside]{article}
\usepackage[utf8]{inputenc} 

\usepackage{asp2014}

\aspSuppressVolSlug
\resetcounters

\begin{document}

\title{Open Science and Artificial Intelligence for supporting the sustainability of the SRC Network: The espSRC case}


\author{Julián~Garrido,$^1$$^,$$^*$ Susana Sánchez-Expósito,$^1$ Antonio~Ruiz-Falcó,$^2$ José~Ruedas,$^1$ María Ángeles Mendoza,$^1$  Victor Vázquez,$^3$ Manuel Parra,$^1$ Jesús Sánchez,$^1$ Ixaka Labadie,$^1$  Laura Darriba,$^3$ Javier Moldón,$^1$ Manuel Rodriguez-Álvarez.,$^3$ Javier Díaz,$^3$$^,$$^4$ and Lourdes Verdes-Montenegro$^1$}
\affil{$^1$Instituto de Astrofísica de Andalucía (IAA), CSIC, Granada, España}
\affil{$^2$Pue Control, Huetor Vega, Spain}
\affil{$^3$Universidad de Granada, Granada, Spain}
\affil{$^4$ Safran Electronics and Defense, Spain}
\affil{$^*$\email{jgarrido@iaa.csic.es}}

\paperauthor{Julián~Garrido}{jgarrido@iaa.csic.es}{0000-0002-6696-4772}{Instituto de Astrofísica de Andalucía - CSIC}{Extagalactic Astronomy}{Granada}{Granada}{18008}{Spain}
\paperauthor{Susana~Sánchez}{sse@iaa.es}{0000-0002-7510-7633}{Instituto de Astrofísica de Andalucía - CSIC}{Extagalactic Astronomy}{Granada}{Granada}{18008}{Spain}
\paperauthor{Lourdes~Verdes-Montenegro}{lourdes@iaa.es}{0000-0003-0156-6180}{Instituto de Astrofísica de Andalucía - CSIC}{Extagalactic Astronomy}{Granada}{Granada}{18008}{Spain}
\paperauthor{Javier Moldón}{jmoldon@iaa.es}{0000-0002-8079-7608}{Instituto de Astrofísica de Andalucía - CSIC}{Extagalactic Astronomy}{Granada}{Granada}{18008}{Spain}
\paperauthor{Laura Darriba}{ldarriba@iaa.es}{0000-0002-5599-2647}{Instituto de Astrofísica de Andalucía - CSIC}{Extagalactic Astronomy}{Granada}{Granada}{18008}{Spain}
\paperauthor{José Ruedas}{jose.ruedas@iaa.csic.es}{0000-0001-6884-8567}{Instituto de Astrofísica de Andalucía - CSIC}{Extagalactic Astronomy}{Granada}{Granada}{18008}{Spain}
\paperauthor{María Ángeles Mendoza}{amendoza@iaa.es}{0000-0003-3800-8668}{Instituto de Astrofísica de Andalucía - CSIC}{Extagalactic Astronomy}{Granada}{Granada}{18008}{Spain}
\paperauthor{Ixaka Labadie}{ixakalab@iaa.es}{0009-0003-3140-7794}{Instituto de Astrofísica de Andalucía - CSIC}{Extagalactic Astronomy}{Granada}{Granada}{18008}{Spain}

\paperauthor{Javier Díaz}{jda@ugr.es}{0000-0002-1849-8068}{Universidad de Granada}{Departamento de Construcciones Arquitectónicas}{Granada}{Granada}{18008}{Spain}
\paperauthor{Manuel Rodríguez}{manolo@ugr.es}{0000-0002-6348-9769}{Universidad de Granada}{Departamento de Construcciones Arquitectónicas}{Granada}{Granada}{18008}{Spain}
\paperauthor{Victor Vázquez}{victorvazrod@ugr.es}{0000-0002-7608-0153}{Universidad de Granada}{Departamento de Construcciones Arquitectónicas}{Granada}{Granada}{18008}{Spain}




\begin{abstract}
The SKA Observatory (SKAO), a landmark project in radio astronomy, seeks to address fundamental questions in astronomy. To process its immense data output, approximately 700 PB/year, a global network of SKA Regional Centres (SRCNet) will provide the infrastructure, tools, computational power needed for scientific analysis and scientific support. The Spanish SRC (espSRC) focuses on ensuring the sustainability of this network by reducing its environmental impact, integrating green practices into data platforms, and developing Open Science technologies to enable reproducible research. This paper discusses and summarizes part of the research and development activities that the team is conducting to reduce the SRC energy consumption at the espSRC and SRCNet. The paper also discusses fundamental research on trusted repositories to support Open Science practices. 
\end{abstract}



\section{Introduction}
The SKA Observatory (SKAO), a key project in radio astronomy, aims to explore essential questions in astrophysics, fundamental physics, and astrobiology. The amount of data generated by the SKA Observatory (SKAO) will reach exa-scale and it will need a Network of SKA Regional Centres (SRCNet, \citet{SRCNetRoadmap2023}) that will serve as a platform to facilitate scientific collaborations as well as access to SKAO data, analysis tools and processing power necessary to fully exploit their science potential. The SRCNet represents the scientific core of the project as it will be the place where the community will analyse SKAO data and will produce advanced data products. 

The SRCNet Resource Board, composed of representatives from the countries participating in SKA, coordinates the allocation of resources required for the SRCNet development project. These resources include personnel, who form international development teams working on the SRCNet implementation using an agile approach.
It is expected that SRCNet supports telescope commissioning and science verification. The Spanish SRC (espSRC) is leading one of the international teams (Coral Team) whose main goal is to build testbeds for evaluating technologies to be considered for the SRCNet platform. Within this context, the espSRC team is conducting research on how to approach the SRCNet development following Green and Open Science practices. 


Section \ref{sect.goals} describes the general goals of this project while section \ref{sect.energy} summarizes our study to understand how to reduce the power associated to cooling in the IAA computing room. Section \ref{sect.Data} explains our approach to reduce data transfer and section \ref{sect.time} explains a development to provide trusted Open Science services. 



\section{Goals and hightlighted activities}\label{sect.goals}

The espSRC (\citet{2023JATISGarrido}) is also leading activities that aim to contribute to create a sustainable SRC Network following Green and Open Science practices. These, are fundamental principles that drive the espSRC goals and activities. In particular, we have the following specific goals: a) minimising the environmental footprint of SRCs; b) incorporating a green perspective into the platforms for SKA data storage, analysis and visualisation within the SRC Network prototypes; and c) developing Open Science technologies supporting end-to-end research reproducibility. These goals are partly addressed during our contribution to the development of the SRCNet, as explained below. 

The espSRC engages in various activities to support the development and sustainability of the SRCNet. A key focus is on creating testbeds to evaluate emerging technologies. This includes deploying a Mini-SRCNet demonstrator, establishing a workflow repository and Virtual Observatory (VO) services to promote Open Science, and integrating the espSRC's storage system into the SKAO Data Lake for efficient data distribution and archiving. These initiatives aim to ensure seamless functionality and accessibility within the SRCNet framework.

Additionally, the espSRC conducts comprehensive network transfer tests to evaluate data transfer capabilities. These tests include profiling and benchmarking astronomy software to optimize its performance and designing benchmarks to assess the computing efficiency of SRC infrastructure. These efforts are critical to meeting the computational demands of SKA operations.

In line with sustainability goals, the espSRC is also developing advanced models to minimize the energy consumption of SRC facilities. This work reflects our commitment to environmental responsibility while maintaining the operational effectiveness of the network.


\section{SRC energy consumption}\label{sect.energy}
The reduction of the carbon footprint of the espSRC will require specific actions and monitoring services to understand the joint impact of the operations and external factors like the temperature. For this reason, we are deploying a network of sensors in the computing center at IAA under the framework of TED4SKA project. The sensors are placed in the racks at the front and rear and in different altitudes. In particular, we have deployed three types of sensors (lh52, lht65n, lht65-e31) whose batteries can last for several years. These three sensors are Long Range LoRaWAN and allow measuring temperature and humidity but lht65-e31 also includes an external probe. Having three types of sensors contributes to the resilience of the system. This distribution and configuration guarantees a good coverage of the room, contributing to a better understanding of the hot and cold areas. In addition, we are currently in the process of deploying power consumption sensors in the computing room. 

The data coming from the sensors are being stored at the espSRC cloud and we have deployed Grafana monitoring tool to facilitate data visualisation  and monitoring of the room. Currently, we are gathering historical data coming from the sensors. In addition, power consumption information is collected by the hypervisors at the espSRC in order to complement the data provided by the power consumption sensors. This data is being used to feed the artificial intelligence models that will allow to predict the thermal behavior of the room, the detection of anomalies and to define rules to reduce its environmental impact. 



\section{Reducing Data transfer}\label{sect.Data}
We have performed various tests to better understand the data transfer and data distribution when using data management tools like Rucio. In addition, we are studying how to contribute to reducing data transfer by moving computation to data or deploying advance services that avoid data movement. e.g. visualisation services. Reducing the requirements to access and analyse the data contribute also to lowering access barriers and therefore to the democratisation of Open Science as well as to the environmental impact. 

We have performed various tests to better understand the data transfer and data distribution when using data management tools like Rucio.
In order to evaluate the performance scalability, files were generated randomly and uploaded to the espSRC Rucio storage. These files were replicated from SPSRC to other Rucio Storage Enterpoints (RSEs). Three files were generated for each size, 200 MB, 500 MB, 1 GB, 1.5 GB, 2 GB and 3GB. The average transfer rate and duration of the three files were uses as metrics for evaluating the performance of RSE and file size.
As expected, the closer the RSE to the espSRC, the higher the transfer rate and lower the transfer duration, so for the Chinese RSE the transfer duration was higher than for Switzerland RSE. As for the size file, the average transfer rate improved for greater files. The impovement was more meaningful for close RSEs.

\section{Timestamps for trusted Open Science}\label{sect.time}

The development of a timestamp service represents a significant advancement in ensuring the traceability and integrity of scientific contributions. By applying timing traceability technologies, the service creates a unique authorship fingerprint, including a code and a traceable timestamp for scientific data. This approach facilitates that researchers are credited for their contributions while promoting the early dissemination of scientific results. The timestamp not only establishes authorship but also safeguards the originality and provenance of research, thereby addressing key challenges in collaborative and data-intensive environments. 

In addition, the provision of timestamps in scientific platforms guarantees the inviolability of the data by creating a unique and verifiable trace of its existence at a specific moment. This process begins with a Time Stamping Request (TSQ), which is generated using tools such as OpenSSL and sent to the timestamp service. The TSQ includes a cryptographic hash of the data being certified, ensuring that the exact content can be authenticated without exposing its details (\citet{Haber1991}). The service, synchronized with a highly accurate time source like ROA (Real Observatorio de la Armada) via NTP, responds with a Time Stamping Reply (TSR), confirming the data existed at the specified time. Both TSQ and TSR are defined using \href{http://www.overleaf.com}{RFC3161} time-stamp protocol. 

The TSR serves as a digital seal that can be incorporated into various formats, as metadata. When integrated into the platform, the timestamp becomes an immutable proof of the data provenance. For verification, the TSR can be read using OpenSSL. Any modification to the file after the timestamp was applied will render the verification invalid, demonstrating tamper-evidence and preserving the data’s integrity. This robust mechanism ensures trust in the originality and authenticity of the research, forming a critical part of Open Science practices and Big Data initiatives like those represented by the SKA.

This timestamp service contributes to advancing Open Science by integrating trusted timestamping into workflows for sharing scientific data and results. Designed with a focus on accessibility, we are exploring the migration to a cloud platform, making it widely available to researchers globally. 

%
%
%
%
%

\section{Conclusions}
This work describes the critical role of the Spanish SKA Regional Centre (espSRC) in advancing sustainable, green, and open practices within the SRCNet framework. By implementing innovative technologies such as timestamp services for trusted Open Science, deploying monitoring systems to optimize energy consumption, and reducing data transfer needs through advanced services, the espSRC addresses both technical and environmental challenges. 

\acknowledgements The Authors, acknowledges financial support from the grant CEX2021-001131-S funded by MICIU/AEI/ 10.13039/501100011033, Ph.D. fellowship program with grant numbers FPU20/05842 and PRE2021-100660, the grants  PID2021-123930OB-C21 and PID2021-123930OB-C22 funded by MICIU/AEI/ 10.13039/501100011033 and by ERDF/EU, the grant TED2021-130231B-I00 funded by MICIU/AEI/ 10.13039/501100011033 and by the European Union NextGenerationEU/PRTR, grant RED2022-134464-T funded by MICIU/AEI/ 10.13039/501100011033, grant INFRA24023 funded by CSIC, and the coordination of the participation in SKA-SPAIN, funded by the Ministry of Science, Innovation and Universities (MICIU). 

\bibliography{P203}  


\end{document}